\documentclass[usenatbib,useAMS,letterpaper]{mn2e}
\usepackage{graphicx}
\usepackage[fleqn]{amsmath}
\usepackage{amssymb}
\usepackage[totalwidth=520pt, totalheight=680pt]{geometry}

\newcommand\apj{ApJ}
\newcommand\apjl{ApJL}
 
\newcommand\mnras{MNRAS} 
\newcommand\aap{A\&A} 
\newcommand\aaps{A\&AS} 
\newcommand\nat{Nature}

\title[No snow-plough for hardening SMBHs binaries]{No snow-plough
  mechanism during the rapid hardening of supermassive black hole
  binaries}

\author[C. Baruteau et al.]{Cl{\'e}ment Baruteau$^{1}$\thanks{E-mail: 
C.Baruteau@damtp.cam.ac.uk}
, Enrico Ramirez-Ruiz$^{2}$ 
and Fr{\'e}d{\'e}ric Masset$^{3}$ 
\\
$^{1}$DAMTP, University of Cambridge, Wilberforce Road, Cambridge CB3 0WA, UK\\
$^{2}$Department of Astronomy and Astrophysics, University of California, Santa Cruz, CA 95064, USA\\
$^{3}$Instituto de Ciencias F{\'i}sicas, UNAM, Apdo. Postal 48-3,
  62251-Cuernavaca, Morelos, M{\'e}xico
}

\begin{document}
\date{Accepted 2012 March 19.  Received 2012 February 21; in original 
form 2012 January 03}
\pagerange{\pageref{firstpage}--\pageref{lastpage}} \pubyear{2011}
\maketitle
\label{firstpage}

\begin{abstract}
  We present two-dimensional hydrodynamical simulations of the tidal
  interaction between a supermassive black hole binary with moderate
  mass ratio, and the fossil gas disc where it is embedded. Our study
  extends previous one-dimensional height-integrated disc models,
  which predicted that the density of the gas disc between the primary
  and the secondary black holes should rise significantly during the
  ultimate stages of the binary's hardening driven by the
  gravitational radiation torque. This snow-plough mechanism, as we
  call it, would lead to an increase in the bolometric luminosity of
  the system prior to the binary merger, which could be detected in
  conjunction with the gravitational wave signal. We argue here that
  the snow-plough mechanism is unlikely to occur.  In two-dimensions,
  when the binary's hardening timescale driven by gravitational
  radiation becomes shorter than the disc's viscous drift timescale,
  fluid elements in the inner disc get funneled to the outer disc
  through horseshoe trajectories with respect to the secondary. Mass
  leakage across the secondary's gap is thus found to be effective
  and, as a result, the predicted accretion disc luminosity will
  remain at roughly the same level prior to merger.
  \end{abstract}


\section{Introduction}
The merger of two galaxies results in the formation of a newly
assembled galaxy containing a supermassive black hole binary in its
nucleus \citep{Begelman80}. Models of hierarchical galaxy formation
predict that the most common supermassive black hole binaries arise
from minor galactic mergers, and have unequal-mass black holes
\citep{Lacey93}. The tidal interaction between the binary black hole,
and the disc of gas and stars it is embedded in, shrinks the binary's
orbit.  The hardening efficiency has a complex
  dependence on the binary's mass ratio \citep{Kazan05}, the gas
  disc-to-binary mass ratio \citep{Lodato09}, and stellar dynamical
  interactions \citep{Milo01}.  Provided the binary hardens to
  sufficiently small separations, the angular momentum extracted from
the binary's orbit becomes progressively more dominated
by gravitational radiation. When the gravitational wave torque
prevails over the disc torque, the disc's accretion rate becomes much
smaller than the binary's hardening rate.

One-dimensional disc models \citep{AN02, Lodato09, Chang10} showed
that the density of the disc region located between the primary and
the secondary black holes should significantly increase prior to
coalescence, as the binary's hardening is dominated by gravitational
radiation. In this scenario, the secondary rapidly drains the inner
disc onto the primary, acting like a snow-plough. The increase in the
inner disc density would lead to a sudden increase in the disc's
bolometric luminosity in the few days prior to merger. This would
constitute an electromagnetic precursor to a supermassive black hole
merger, which could be identified in conjunction with the
gravitational wave signal detectable with the Laser Interferometer
Space Antenna (LISA) \citep{Chang10}.

In this Letter, we show that this prediction does not hold in
two-dimensional (2D) disc models. We demonstrate with the help of 2D
hydrodynamical simulations that when the binary's hardening is
dominated by gravitational radiation, the inner disc is progressively
funneled to the outer disc through horseshoe trajectories with respect
to the secondary, with the consequence that there should be no
significant increase in the disc luminosity prior to the merger.  The
physical problem we consider is described in \S~\ref{sec:pb}, and the
numerical setup of our simulations is detailed in
\S~\ref{sec:setup}. Our results of calculations are presented in
\S~\ref{sec:results}, followed by some concluding remarks in
\S~\ref{sec:cl}.

\section{Physical problem}
\label{sec:pb}
We examine the tidal interaction between a supermassive black hole
binary with mass ratio $q \ll 1$ and the gaseous disc it is embedded
in, neglecting for simplicity the presence of other stars. The torque
responsible for the binary's hardening comprises the tidal torque
$\Gamma_{\rm disc}$ exerted by the gaseous disc, and the torque
$\Gamma_{\rm GW}$ due to the emission of gravitational waves (GW).
The latter reads \citep[e.g.,][]{AN05}
\begin{equation}
  \Gamma_{\rm GW} = -\frac{32}{5}\frac{G^3}{c^5 a^3}\left[\frac{G(M_1+M_2)}{a}\right]^{1/2}M_1(M_1+M_2)M_2^2
\label{gammagw}
\end{equation}
for a circular binary, where $M_1$ and $M_2 = qM_1$ denote the mass of
the primary and of the secondary, respectively, $a$ is the binary's
semi-major axis, $c$ is the speed of light, and $G$ is the
gravitational constant. Since $|\Gamma_{\rm GW}|$ strongly increases
with decreasing $a$, the binary's hardening is ultimately driven by
the gravitational wave torque. The binary's hardening timescale
$\tau_{\rm GW}$ due to the GW torque is
\begin{equation}
  \tau_{\rm GW} = \frac{5}{256} \frac{c^5 a^4}{G^3} M_1^{-1}M_2^{-1}(M_1+M_2)^{-1}.
\label{taugw1}
\end{equation}
Further denoting the Schwarzschild radius $r_{\rm g} = 2GM_1 / c^2$,
Eq.~(\ref{taugw1}) can be recast as
\begin{equation}
  \frac{\tau_{\rm GW}}{T_{\rm orb}} \approx \frac{1.76\times 10^{-2}}{q\sqrt{1+q}} \left( \frac{a}{r_{\rm g}}\right)^{5/2},
\label{taugw}
\end{equation}
where $T_{\rm orb}$ is the orbital period at the binary's semi-major
axis $a$. At the innermost stable circular orbit (ISCO), located at
$r_{\rm isco} = 3r_{\rm g}$ for a non-rotating primary black hole,
$\tau_{\rm GW} / T_{\rm orb} \approx 2.4, 25$ and 250 for $q=10^{-1}$,
$10^{-2}$ and $10^{-3}$, respectively.

In this study, the secondary to primary mass ratio is fixed to
$q=2\times 10^{-2}$, and this large value implies that the secondary
is able to open a gap around its orbit even in a relatively thick and
viscous disc \citep{lp86, crida06}. It also makes the two-dimensional
approximation for the disc applicable when discarding gas accretion
onto the secondary \citep{gda2005}. We point out, as did
\cite{Gould00}, that apart from the presence of the GW torque, the
physical problem we consider shares a number of analogies with the
orbital evolution of a massive gap-opening planet in a protoplanetary
disc.

We examine the evolution of the fluid elements in the inner disc close
to the location of the inner separatrix of the secondary's horseshoe
region. These fluid elements are on approximately circulating
streamlines with respect to the secondary, with a relative (synodic)
period $\tau_{\rm syn} \sim$ half the horseshoe libration period. The
horseshoe libration period reads $\tau_{\rm lib} = 8\pi a \times (3
\Omega x_{\rm s})^{-1}$, where $\Omega$ is the secondary's angular
frequency, and $x_{\rm s}$ is the radial half-width of the horseshoe
region.  For gap-opening secondaries, $x_{\rm s} \sim 2R_{\rm H}$
\citep{mak2006}, where $R_{\rm H} = a (q/3)^{1/3}$ is the Hill radius
of the secondary.
 
During a synodic period, if the radial distance $\delta a_{\rm b}$ by
which the binary hardens becomes comparable to, or greater than the
radial distance $\delta a_{\rm s}$ by which fluid elements near the
inner separatrix drift inward, these fluid elements embark on
horseshoe trajectories and are funneled to the outer disc. The value
of $\delta a_{\rm s}$ is essentially set by the viscous torque, and
$\delta a_{\rm b}$ by $\Gamma_{\rm disc} + \Gamma_{\rm GW} \approx
\Gamma_{\rm GW}$ when the binary shrinkage is dominated by
gravitational radiation. Thus, in order of magnitude, $\delta a_{\rm
  b} \gtrsim \delta a_{\rm s}$ when $\tau_{\rm GW}$ becomes shorter
than the viscous drift timescale $\sim 2a^2 / 3\nu$, where $\nu =
\alpha h^2 a^2 \Omega_{\rm K}$ is the disc's turbulent viscosity
($\alpha$ denotes the alpha viscous parameter, h the disc's aspect
ratio, and $\Omega_{\rm K}$ the Keplerian angular velocity). Using
Eq.~(\ref{taugw}), the typical binary separation below which $\delta
a_{\rm b} \gtrsim \delta a_{\rm s}$, often called
  decoupling radius, can thus be estimated as
\begin{equation}
  \frac{a}{r_{\rm g}} \sim 13 \times 
  \left(\frac{\alpha}{0.03} \right)^{-2/5} 
  \left(\frac{q}{0.02} \right) ^{2/5} 
  \left(\frac{h}{0.08} \right)^{-4/5},
  \label{acrit}
\end{equation}
where $\alpha$ and $h$ are to be evaluated at the secondary's
location.  Detailed modeling of the disc's thermal
  balance yields a decoupling radius that is a few times larger than
  that given in Eq.~(\ref{acrit}) \citep{Milo05}. This illustrates
  that when the hardening of an unequal-mass binary black hole is
dominated by gravitational radiation, a substantial funneling of the
inner disc to the outer disc is possible beyond the ISCO location,
which we confirm with 2D hydrodynamical simulations in
\S~\ref{sec:results}.

\section{Physical model and numerical setup}
\label{sec:setup}
We investigate the tidal interaction between a black hole binary and
the gas disc it is embedded in, assuming the binary's hardening is
dominated by the emission of gravitational waves. For this purpose, 2D
hydrodynamical simulations were carried out with the {\sc Fargo}
code \citep[][\texttt{http://fargo.in2p3.fr}]{fargo1}. A
cylindrical coordinate system $\{r,\varphi\}$ centred onto the primary
black hole is adopted.  The disc extends from $r=1.2 r_{\rm g}$ to
$r=35.4r_{\rm g}$ (that is, $r \in [0.4-11.8] r_{\rm isco}$ with
assuming a non-rotating primary).
\\
\par\noindent\emph{Binary parameters---}
The mass of the primary and secondary black holes is $M_1 =
5\times10^8 M_{\odot}$ and $M_2 = qM_1 = 10^7 M_{\odot}$,
respectively, as in \cite{AN02}.  The initial semi-major axis of the
binary is $a_0 \approx 11.8 r_{\rm g} \approx 5.7 \times 10^{-4}$ pc
(or, $a_0 \sim 4r_{\rm isco} $). The binary is assumed to be circular,
and its orbital plane to be aligned with the disc. Each black hole is
treated as a point mass potential. Relativistic effects are discarded
for simplicity and to facilitate comparison with previous studies.
Similarly, gas accretion onto the secondary is neglected. Also, to
avoid a large accumulation of mass inside its Hill radius, the
secondary's gravitational potential is smoothed over a softening
length $\varepsilon = H(a_0)$, where $H$ is the pressure scale height
(its value is specified below).
\\
\par\noindent\emph{Disc parameters---}
The disc is initially axisymmetric and rotates at the angular
frequency $\Omega(r)$ about the primary. The expression for
$\Omega(r)$ assumes a radial equilibrium between the centrifugal
acceleration, the radial acceleration due to the pressure gradient,
and the gravitational acceleration due to the primary alone. The
initial gas density is $\approx 10^7$ g\,cm$^{-2} \times (r /
a_0)^{-1/2}$, which corresponds to an initial mass $\approx 6.5\times
10^{-4}\,M_1$. For simplicity, a locally isothermal equation of state
is considered where the initial radial profile of the disc
temperature, which we take proportional to $r^{-3/2}$, remains
stationary. The disc temperature is conveniently related to the disc's
aspect ratio $h = H/r$, which we take to be $h(r) = 0.08 \times
(r/a_0)^{-1/4}$. The disc turbulence is modeled with a constant
kinematic viscosity $\nu$ that corresponds to an alpha parameter
$\alpha \approx 0.03$ uniformly throughout the disc.
\\
\par\noindent\emph{Numerical setup---}
The grid has $N_r = 280$ cells along the radial direction, and it
covers the full $2\pi$ range in azimuth with $N_{\varphi}=780$ cells.
A logarithmic spacing is used along the radial direction to optimize
the grid's resolution near the disc's inner edge.  The secondary's
Hill radius is resolved by about 30 grid cells along each direction at
all times.  Standard zero-gradient outflow boundary conditions are
applied at the grid's inner and outer edges.
\begin{figure}
  \resizebox{\hsize}{!}{\includegraphics{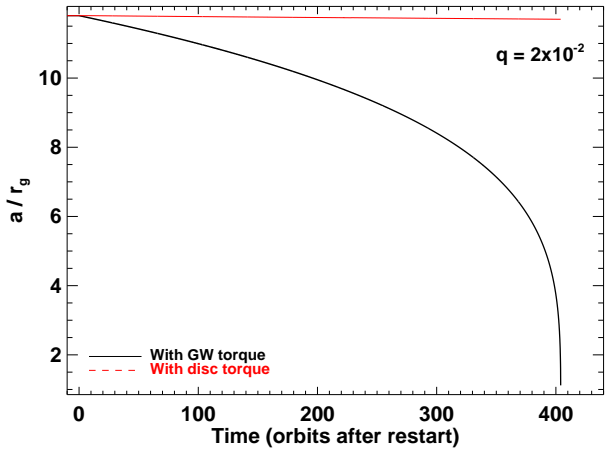}}
  \caption{\label{fig:fig1} Time evolution of the binary's semi-major
    axis, $a$, driven by the GW torque alone (black curve), and by the
    disc torque alone (red curve, both are results of simulations).
    Time is expressed in units of the binary's orbital period at its
    initial separation, and $a$ is shown in units of $r_{\rm g}$.}
\end{figure}
\begin{figure*}
  \centering\resizebox{\hsize}{!}
  {
   \includegraphics{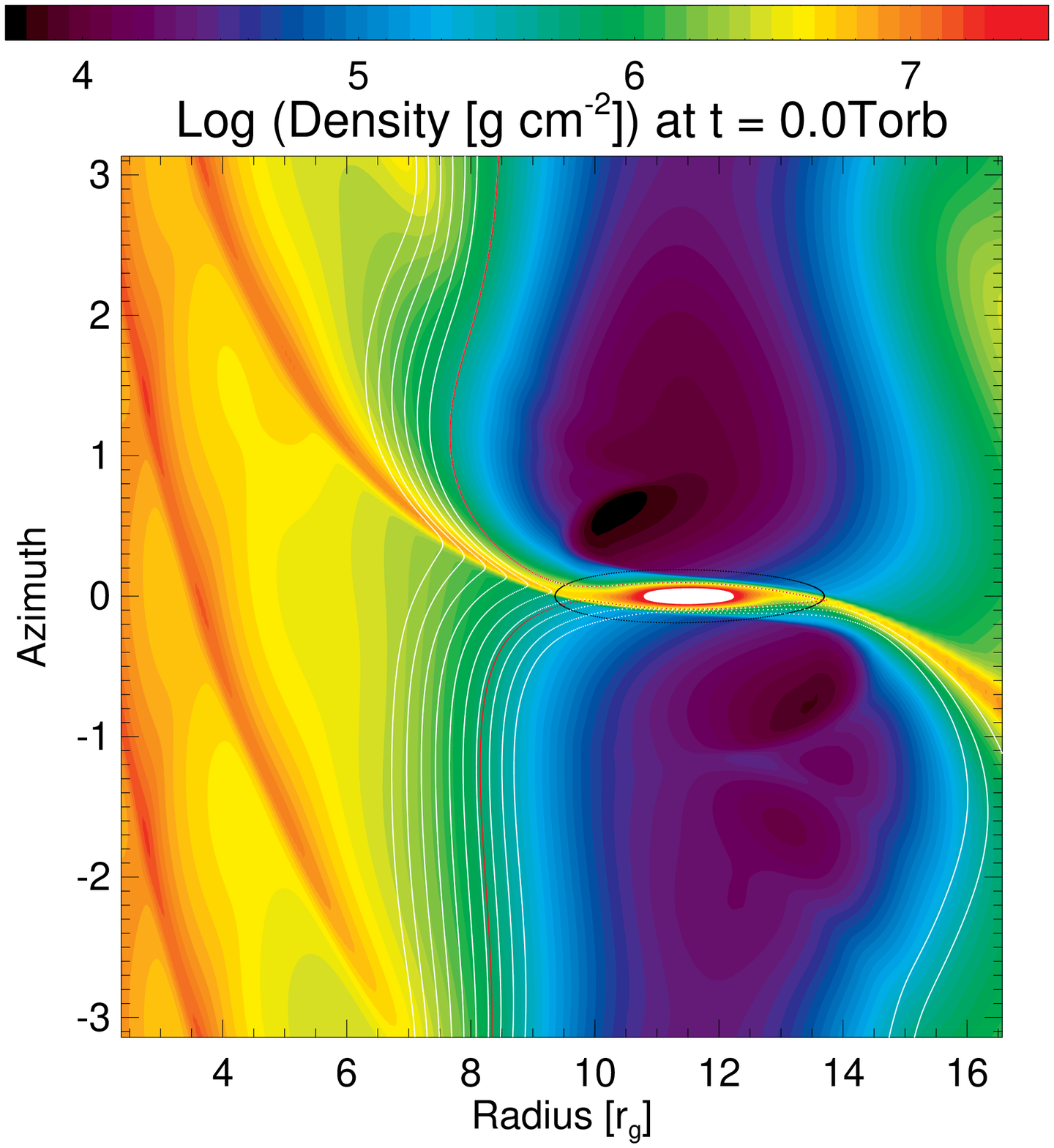}
   \includegraphics{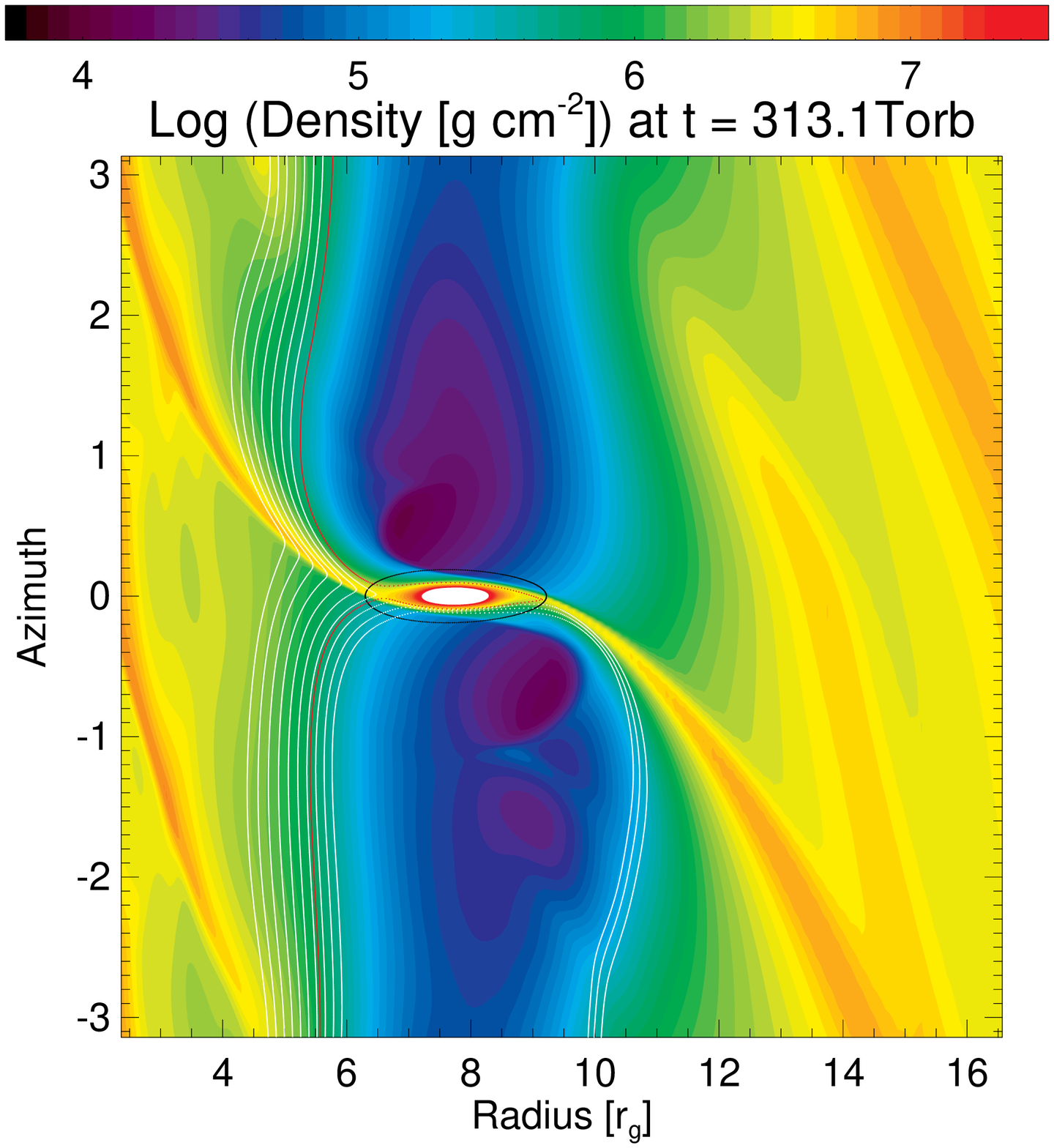}
   \includegraphics{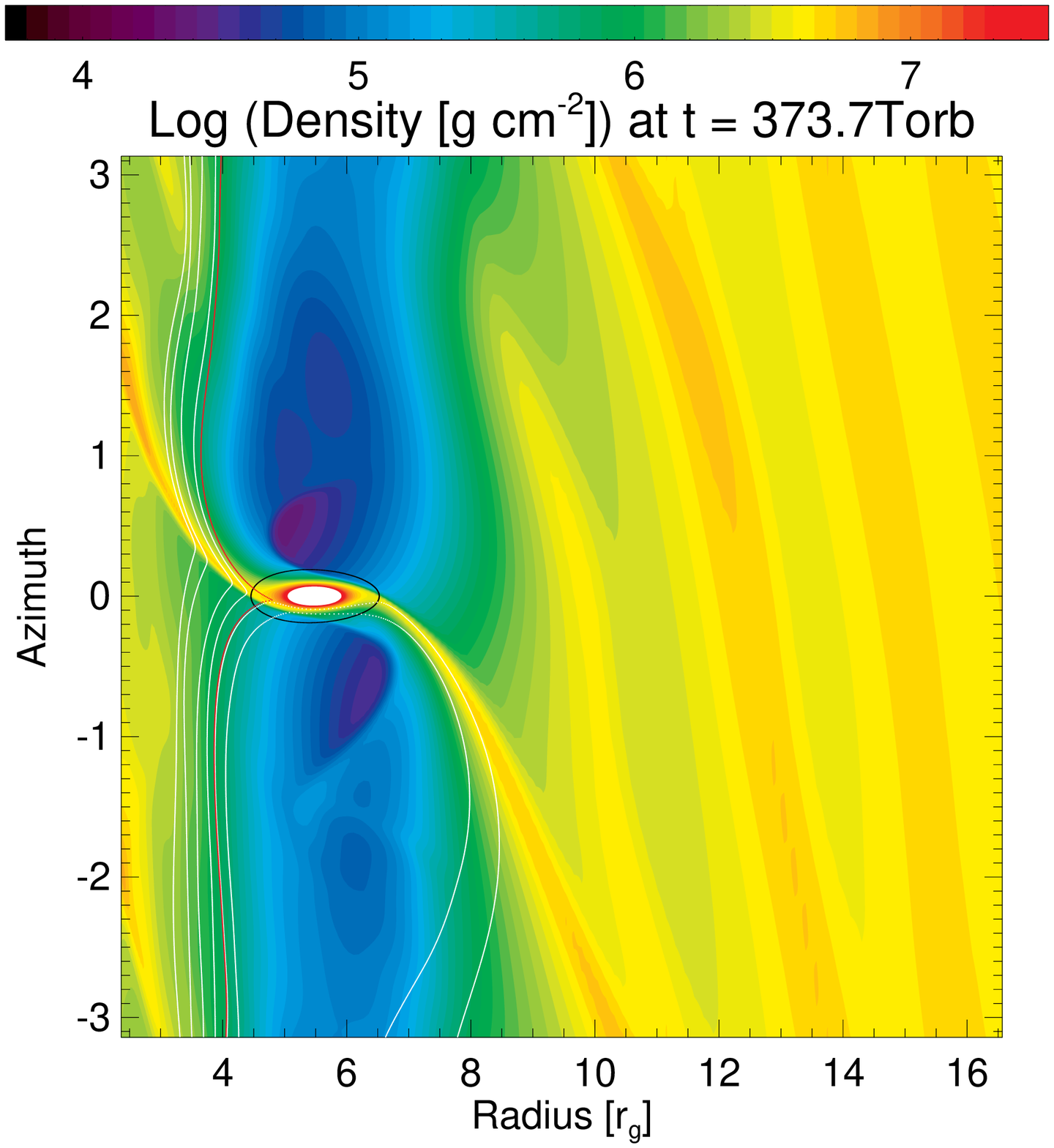}
  }
  \caption{\label{fig:fig2} Time sequence of the gas surface density
    when the binary's hardening is driven by the GW torque.  The
    $x-$axis shows the radial coordinate in units of the Schwarzschild
    radius $r_{\rm g}$ (part of the grid is shown along this
    direction), and the $y-$axis shows the azimuth.  The same color
    scale is adopted in all three panels. Time is in units of the
    binary's orbital period before the hardening stage. The binary's
    Hill radius is highlighted by a black circle, and a few
    streamlines in the frame rotating with the secondary are
    overplotted by white curves.  The location of the circular
    streamline of the inner disc closest to the secondary is shown by
    a red curve.}
\end{figure*}

\section{Results}
\label{sec:results}
Our simulations were performed in two steps. The secondary is first
held on a fixed circular orbit about the binary's center of mass. Its
mass is gradually increased over 100 orbits to avoid a violent disc
relaxation following the introduction of the secondary. The secondary
progressively depletes a gap around its orbit, which attains a quasi
steady-state density profile $\sim$ 400 orbits after the insertion of
the secondary. Simulations were then restarted with the binary feeling
the gravitational wave torque only. This second stage, which we refer
to as the hardening stage in what follows, lasts $\lesssim 410$
orbits, as indicated by Eq.~(\ref{taugw}). The disc torque on the
secondary is discarded for simplicity, as $|\Gamma_{\rm disc}| \ll
|\Gamma_{\rm GW}|$ from the binary's initial separation. This point is
clearly illustrated in Fig.~\ref{fig:fig1}, where we compare the time
evolution of the binary's semi-major axis when only the GW torque is
included (black curve), and when only the disc torque is included (red
curve).

Fig.~\ref{fig:fig2} shows a time sequence of the gas surface density
during the hardening stage. Time is expressed in orbital periods at
the binary's initial separation, $a_0$. Streamlines in the frame
rotating with the secondary are shown as white curves. The circulating
streamline of the inner disc closest to the secondary is depicted by a
red curve. This time sequence clearly shows no density increase in the
inner disc, but underlines instead the progressive replenishment of
the gap. The gas density inside the gap becomes more and more
asymmetric as the binary's hardening gets faster, with more gas behind
the secondary ($\varphi < 0$) than ahead of it.  This asymmetry
results from fluid elements of the inner disc embarking onto horseshoe
streamlines and being funneled to the outer disc or to the gap. This
gap asymmetry is reminiscent of the one occurring during type III
runaway migration \citep[a fast, generally inward migration regime
driven by the disc torque; see][]{mp03}.

\begin{figure}
  \resizebox{\hsize}{!}
  {
    \includegraphics[width=\hsize]{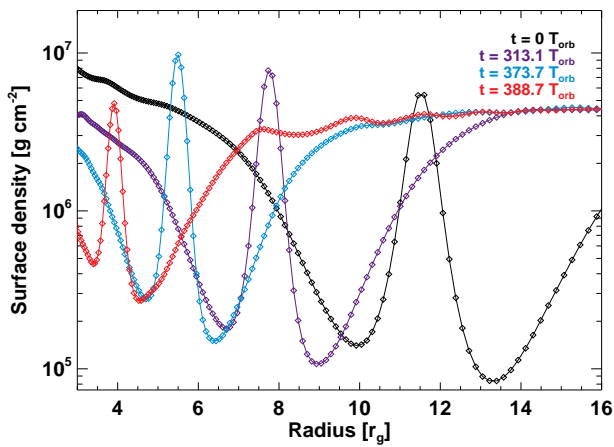}
  }
  \caption{\label{fig:fig3} Azimuthally-averaged surface density at
    different times expressed in units of the binary's initial orbital
    period. Radius is in units of the Schwarzschild radius $r_{\rm
      g}$. Density peaks correspond to gas accumulation
      inside the circum-secondary disc.}
\end{figure}

Fig.~\ref{fig:fig3} displays the azimuthally-averaged disc density at
different times prior to the binary's merger. It further illustrates
that the rapid shrinkage of the binary does not squeeze the inner disc
in our 2D simulations, in contrast to previous height-integrated 1D
disc models \citep{AN02, Lodato09, Chang10}.

\begin{figure*}
  \centering\resizebox{\hsize}{!}
  {
   \includegraphics{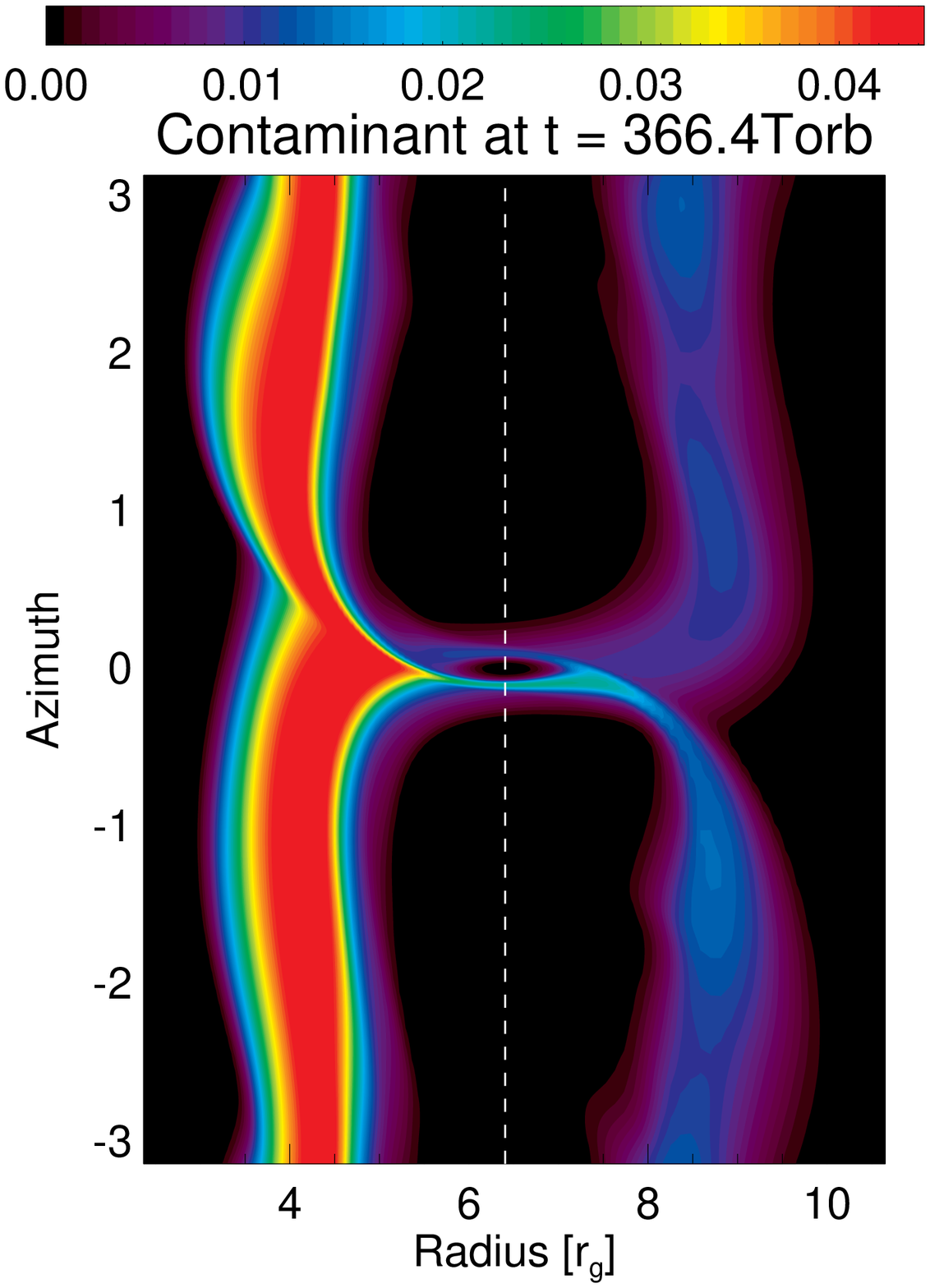}
   \includegraphics{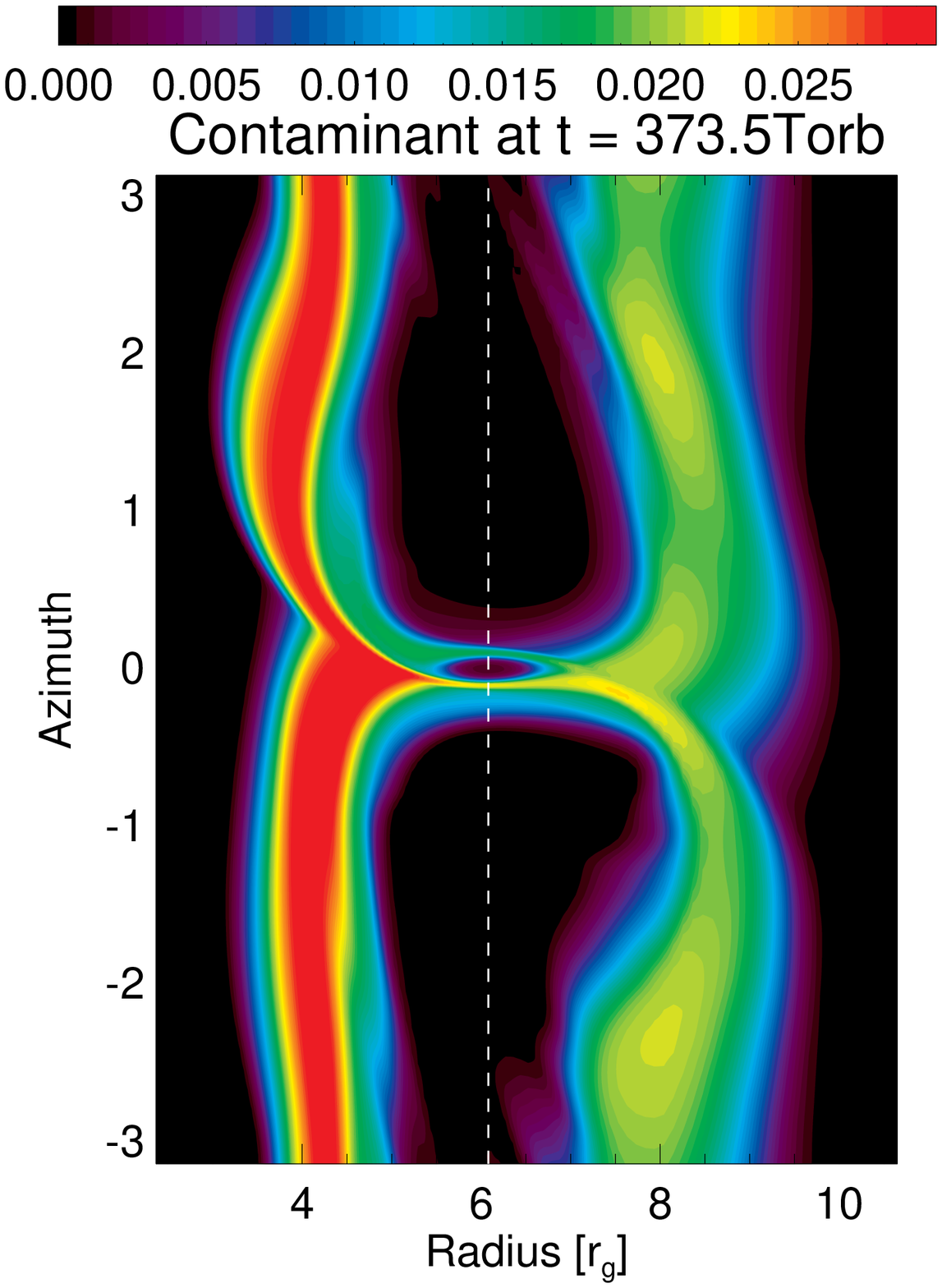}
   \includegraphics{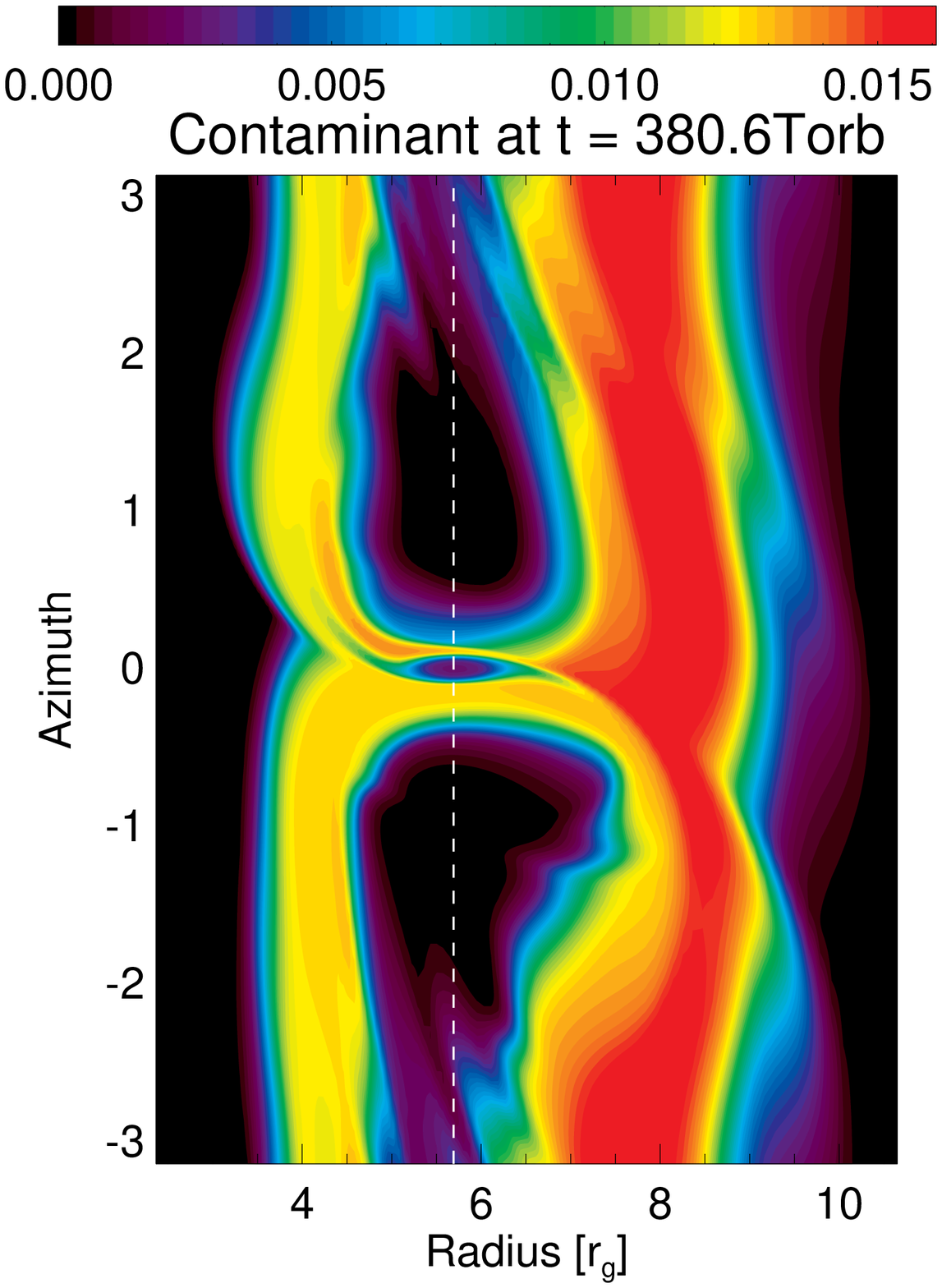}
   \includegraphics{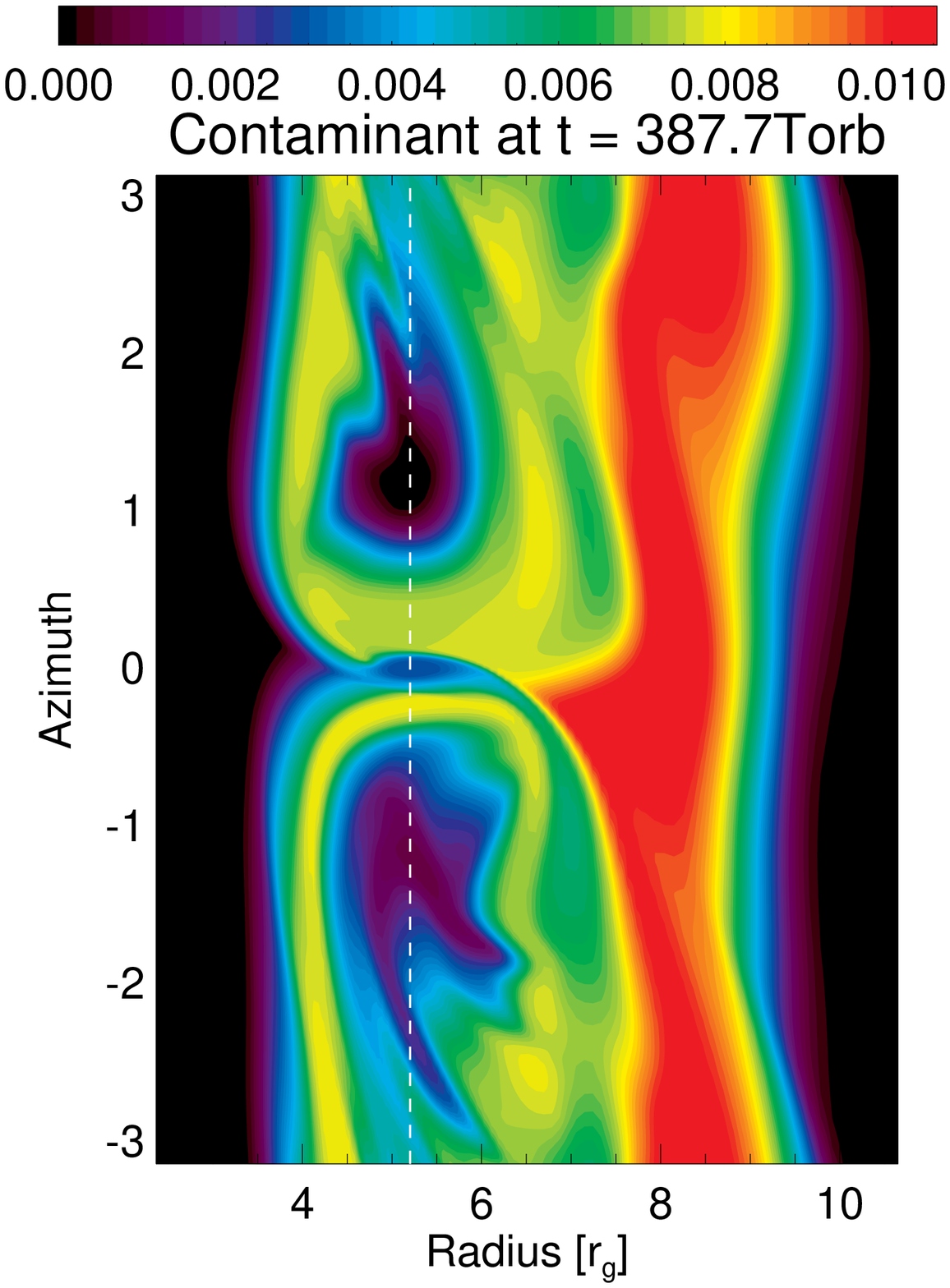}
  }
  \caption{\label{fig:fig4}Time sequence showing the evolution of the
    specific concentration of the passive contaminant, $C$, defined in
    the text. The $x-$axis shows the radial coordinate in units of
    $r_{\rm g}$, and the $y-$axis the azimuth. Contour levels are
    defined for each panel to underline the region occupied by the
    contaminant at each time. The binary's separation is indicated by
    a vertical dashed line.}
\end{figure*}
To gain further insight into the time evolution of the mass in the
inner disc, we have performed simulations in which a small patch of
the disc inside the inner separatrix of the horseshoe region is
polluted initially with a passive contaminant that is advected with
the flow. We denote by $\psi$ the concentration per unit area of the
contaminant, and evolve the equation
\begin{equation}
  \frac{\partial \psi}{\partial t} + \nabla \cdot (\psi {\bf v}) = 0.
\label{eqn:scalar}
\end{equation}
Writing $\psi(r, \varphi) = C(r, \varphi) \, \Sigma$, where $C(r,
\varphi)$ is referred to as the specific concentration of the
contaminant, and using the continuity equation, Eq.~(\ref{eqn:scalar})
can be recast as $\partial_t C +{\bf v}\cdot \nabla C = 0$.  The patch
of contaminant is introduced 340 orbits after the beginning of the
hardening stage, and the time evolution of its specific concentration,
$C$, is displayed as a time sequence in Fig.~\ref{fig:fig4}. Time
increases by $\approx 7$ orbits moving from left to right through the
panels.  The location of the secondary is shown by a vertical dashed
line in each panel. In the left-hand panel, most of the contaminant is
concentrated inside the inner separatrix of the horseshoe region,
located at $r \approx 4.5 r_{\rm g}$. At this time, a (relatively)
tiny fraction of the contaminant is being funneled to the outer disc,
and ends up sliding along the outer separatrix of the horseshoe region
at $r \approx 8 r_{\rm g}$. The second panel highlights that more and
more contaminant is being redirected to the outer disc. As the binary
continues hardening, most of the contaminant now moving in the outer
disc remains in the outer disc, only a small fraction of it undergoes
inward horseshoe U-turns back to the inner disc. In the third panel,
most of the contaminant is now concentrated in the outer disc. The
radial profile of the contaminant gets progressively thicker in the
outer disc, because the outer separatrix of the horseshoe region is
moving inwards faster than the contaminant may drift inward
viscously. This is even more clear in the fourth panel of
Fig.~\ref{fig:fig4}, where the contaminant in the outer disc is left
well behind the secondary.  We also mention that in addition to the
material passing to the outer disc, we never observe the contaminant
being pushed inward: the region below $r \sim 3.5 r_{\rm g}$ remains
dark.

\begin{figure}
  \resizebox{\hsize}{!}
  {
    \includegraphics[width=\hsize]{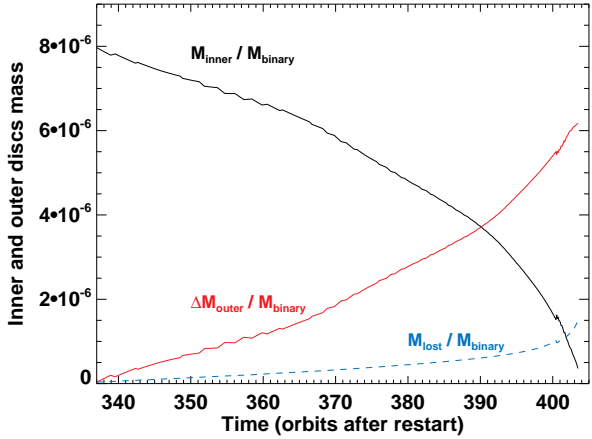}
  }
  \caption{\label{fig:fig5}Mass of the inner disc during the ultimate
    stages of the binary's hardening, before the secondary leaves the
    computational grid (black curve).  The increase in the mass of the
    outer disc during the same time interval, namely $\Delta M_{\rm
      outer} = M_{\rm outer} (t) - M_{\rm outer}(337T_{\rm orb})$,
    with $M_{\rm outer}$ the outer disc's mass, is displayed by a red
    curve. The mass lost from the computational grid during the same
    time interval is shown by a dashed curve.  Masses are in units of
    the binary mass, and all curves have been smoothed over 5 orbits.}
\end{figure}
The time evolution of the gas surface density in Figs.~\ref{fig:fig2}
and~\ref{fig:fig3}, and of the contaminant's specific concentration in
Fig.~\ref{fig:fig4}, suggest that most of the mass in the inner disc
relative to the secondary is transferred to the outer disc through
horseshoe streamlines. To help quantify this funneling mechanism, we
display in Fig.~\ref{fig:fig5} the time evolution of the mass in the
inner disc, in units of the binary mass, during the last $\sim 70$
orbits before the binary's merger.  The increase in the mass of the
outer disc during the same time span (also in units of the binary
mass) is overplotted by a red curve. The difference between both
curves accounts for the mass lost from the computational grid, which
is shown by a dashed curve in Fig.~\ref{fig:fig5}. These curves
clearly show that the fraction of the inner disc's mass funneled to
the outer disc largely exceeds that drained onto the primary.

\section{Concluding remarks}
\label{sec:cl}
The minor merger of two galaxies leads to the formation of a
supermassive black hole binary with unequal mass ratio embedded in a
gaseous disc. The tidal interaction between the binary and the disc
hardens the binary's orbit. When the binary gets sufficiently tight,
emission of gravitational waves becomes the main source of angular
momentum extraction from the binary's orbit, and causes further rapid
shrinkage until coalescence takes place.

We have focused in this Letter on the evolution of the disc region
located between the primary and the secondary black holes (the inner
disc), when the binary's hardening is dominated by the emission of
gravitational waves. With the help of 2D hydrodynamical simulations,
we have shown that the rapid hardening of the binary does not lead to
a significant squeezing of the inner disc. The latter is redirected
instead toward the disc region beyond the secondary's orbit (the outer
disc) through horseshoe streamlines. When the binary's hardening
timescale driven by gravitational radiation becomes shorter than the
disc's viscous drift timescale, fluid elements in the inner disc
embark on horseshoe trajectories with respect to the secondary, and
are progressively funneled to the outer disc.  The funneling of the
inner disc toward the outer disc implies that, in contrast to the
predictions of 1D disc models, the accretion rate onto the primary is
not dramatically increased just prior to merger, and, as a result, the
disc emission before the binary merger should remain at about the same
level. After the merger, an electromagnetic afterglow could be
detected as the disc gets accreted by the merged black hole
\citep{Milo05}.

The physical model we have considered is a straightforward 2D
extension of the model considered in \cite{AN02}, \cite{Lodato09}, and
\cite{Chang10}, and as was already pointed out by these authors, this
model has a number of simplifying assumptions.  We have considered an
intermediate binary mass ratio ($q=2\times 10^{-2}$). A smaller mass
ratio would decrease the binary's separation below which a significant
funneling occurs (which we have checked with additional simulations,
not reported here). A mass ratio closer to unity would initially lead
to a rapid depletion of the inner disc, and the binary would be
surrounded by a circumbinary disc \citep[e.g.,][]{MFM08, Cuadra09},
making our funneling mechanism not applicable. Also, different values
of the disc aspect ratio and viscosity would change the structure of
the gap opened by the secondary (width, depth). Partial
  gap-opening could occur in very thick discs, and a three-dimensional
  disc modeling would be valuable. In some cases, the
disc--secondary interaction can make the outer disc eccentric, thereby
driving the binary's eccentricity \citep[e.g.,][]{Papaloizou01,
  og2003}, although the latter should be partially damped during the
fast inspiral driven by gravitational radiation \citep{AN05}. Still it
remains to be clarified how the funneling mechanism would operate in
the presence of an eccentric binary.

\section*{Acknowledgments}
We thank P. Chang, J. Guilet, Z. Haiman, D. N. C. Lin, K. Menou, and
E. Quataert for stimulating discussions, and the referee for useful
comments.  CB is supported by a Herchel Smith Postdoctoral Fellowship.
ER-R acknowledges support from the David and Lucille Packard
Foundation and the NSF grant: AST-0847563.

\bibliographystyle{mn2e}

\begin{thebibliography}{}

\bibitem[\protect\citeauthoryear{{Armitage} \& {Natarajan}}{{Armitage} \&
  {Natarajan}}{2002}]{AN02}
{Armitage} P.~J.,  {Natarajan} P.,  2002, \apjl, 567, L9

\bibitem[\protect\citeauthoryear{{Armitage} \& {Natarajan}}{{Armitage} \&
  {Natarajan}}{2005}]{AN05}
{Armitage} P.~J.,  {Natarajan} P.,  2005, \apj, 634, 921

\bibitem[\protect\citeauthoryear{{Begelman}, {Blandford} \& {Rees}}{{Begelman}
  et~al.}{1980}]{Begelman80}
{Begelman} M.~C.,  {Blandford} R.~D.,    {Rees} M.~J.,  1980, \nat, 287, 307

\bibitem[\protect\citeauthoryear{{Chang}, {Strubbe}, {Menou} \&
  {Quataert}}{{Chang} et~al.}{2010}]{Chang10}
{Chang} P.,  {Strubbe} L.~E.,  {Menou} K.,    {Quataert} E.,  2010, \mnras,
  407, 2007

\bibitem[\protect\citeauthoryear{{Crida}, {Morbidelli} \& {Masset}}{{Crida}
  et~al.}{2006}]{crida06}
{Crida} A.,  {Morbidelli} A.,    {Masset} F.,  2006, Icarus, 181, 587

\bibitem[\protect\citeauthoryear{{Cuadra}, {Armitage}, {Alexander} \&
  {Begelman}}{{Cuadra} et~al.}{2009}]{Cuadra09}
{Cuadra} J.,  {Armitage} P.~J.,  {Alexander} R.~D.,    {Begelman} M.~C.,  2009,
  \mnras, 393, 1423

\bibitem[\protect\citeauthoryear{{D'Angelo}, {Bate} \& {Lubow}}{{D'Angelo}
  et~al.}{2005}]{gda2005}
{D'Angelo} G.,  {Bate} M.~R.,    {Lubow} S.~H.,  2005, \mnras, 358, 316

\bibitem[\protect\citeauthoryear{{Gould} \& {Rix}}{{Gould} \&
  {Rix}}{2000}]{Gould00}
{Gould} A.,  {Rix} H.-W.,  2000, \apjl, 532, L29

\bibitem[\protect\citeauthoryear{{Kazantzidis}, {Mayer}, {Colpi}, {Madau},
  {Debattista}, {Wadsley}, {Stadel}, {Quinn} \& {Moore}}{{Kazantzidis}
  et~al.}{2005}]{Kazan05}
{Kazantzidis} S.,  {Mayer} L.,  {Colpi} M.,  {Madau} P.,  {Debattista} V.~P.,
  {Wadsley} J.,  {Stadel} J.,  {Quinn} T.,    {Moore} B.,  2005, \apjl, 623,
  L67

\bibitem[\protect\citeauthoryear{{Lacey} \& {Cole}}{{Lacey} \&
  {Cole}}{1993}]{Lacey93}
{Lacey} C.,  {Cole} S.,  1993, \mnras, 262, 627

\bibitem[\protect\citeauthoryear{{Lin} \& {Papaloizou}}{{Lin} \&
  {Papaloizou}}{1986}]{lp86}
{Lin} D.~N.~C.,  {Papaloizou} J.,  1986, \apj, 309, 846

\bibitem[\protect\citeauthoryear{{Lodato}, {Nayakshin}, {King} \&
  {Pringle}}{{Lodato} et~al.}{2009}]{Lodato09}
{Lodato} G.,  {Nayakshin} S.,  {King} A.~R.,    {Pringle} J.~E.,  2009, \mnras,
  398, 1392

\bibitem[\protect\citeauthoryear{{MacFadyen} \&
  {Milosavljevi{\'c}}}{{MacFadyen} \& {Milosavljevi{\'c}}}{2008}]{MFM08}
{MacFadyen} A.~I.,  {Milosavljevi{\'c}} M.,  2008, \apj, 672, 83

\bibitem[\protect\citeauthoryear{{Masset}}{{Masset}}{2000}]{fargo1}
{Masset} F.,  2000, \aaps, 141, 165

\bibitem[\protect\citeauthoryear{{Masset}, {D'Angelo} \& {Kley}}{{Masset}
  et~al.}{2006}]{mak2006}
{Masset} F.~S.,  {D'Angelo} G.,    {Kley} W.,  2006, \apj, 652, 730

\bibitem[\protect\citeauthoryear{{Masset} \& {Papaloizou}}{{Masset} \&
  {Papaloizou}}{2003}]{mp03}
{Masset} F.~S.,  {Papaloizou} J.~C.~B.,  2003, \apj, 588, 494

\bibitem[\protect\citeauthoryear{{Milosavljevi{\'c}} \&
  {Merritt}}{{Milosavljevi{\'c}} \& {Merritt}}{2001}]{Milo01}
{Milosavljevi{\'c}} M.,  {Merritt} D.,  2001, \apj, 563, 34

\bibitem[\protect\citeauthoryear{{Milosavljevi{\'c}} \&
  {Phinney}}{{Milosavljevi{\'c}} \& {Phinney}}{2005}]{Milo05}
{Milosavljevi{\'c}} M.,  {Phinney} E.~S.,  2005, \apjl, 622, L93

\bibitem[\protect\citeauthoryear{{Ogilvie} \& {Lubow}}{{Ogilvie} \&
  {Lubow}}{2003}]{og2003}
{Ogilvie} G.~I.,  {Lubow} S.~H.,  2003, \apj, 587, 398

\bibitem[\protect\citeauthoryear{{Papaloizou}, {Nelson} \&
  {Masset}}{{Papaloizou} et~al.}{2001}]{Papaloizou01}
{Papaloizou} J.~C.~B.,  {Nelson} R.~P.,    {Masset} F.,  2001, \aap, 366, 263

\end{thebibliography}

\label{lastpage}
\end{document}